\documentclass{article}
\usepackage{spconf}

\usepackage{graphicx}
\usepackage{amsmath}
\usepackage[psamsfonts]{amssymb}
\usepackage{amsfonts}
\usepackage{amsxtra}
\usepackage{mathrsfs}
\usepackage{threeparttable}
\usepackage{arydshln}
\usepackage{cite}
\usepackage{url}
\usepackage{multirow}
\usepackage{enumitem}

\makeatletter
\let\MYcaption\@makecaption
\makeatother
\usepackage{subcaption}
\makeatletter
\let\@makecaption\MYcaption
\makeatother
\makeatletter
\renewcommand\section{\@startsection{section}{1}{\z@}
                      {0.5ex \@plus 0ex \@minus -1ex}
                      {0.5ex \@plus 0ex}
                      {\normalfont\Large\bfseries}}
\renewcommand\subsection{\@startsection{subsection}{2}{\z@}
                      {0.5ex \@plus 0ex \@minus -1ex}
                      {0.5ex \@plus 0ex}
                      {\normalfont\large\bfseries}}
\renewcommand\subsubsection{\@startsection{subsubsection}{3}{\z@}
                      {0.5ex \@plus 0ex \@minus -1ex}
                      {0.5ex \@plus 0ex}
                      {\normalfont\normalsize\bfseries}}
\def\@listi{\leftmargin\leftmargini
            \parsep 1.0pt
            \topsep 0.2\baselineskip \@minus 0.1\baselineskip
            \itemsep 1.0pt \relax}
\let\@listI\@listi
\makeatother

\newcommand{\trace}[1]{\mathrm{tr}(#1)}
\newcommand{\diag}[1]{\mathrm{diag}(#1)}
\newcommand{\myvector}[1]{\boldsymbol{#1}}
\newcommand{\mymatrix}[1]{\boldsymbol{\mathrm{#1}}}
\newcommand{\myset}[1]{\mathrm{#1}}
\newcommand{\gaussian}[3]{\mathcal{N}(#1 | #2, #3)}
\newcommand{\prob}[1]{p(#1)}
\DeclareMathOperator*{\argmin}{arg\,min}

\newcounter{num}

\title{Separated-Spectral-Distribution Estimation \\
  Based on Bayesian Inference with Single RGB Camera}
\name{Yuma Kinoshita and Hitoshi Kiya}
\address{Tokyo Metropolitan University, Tokyo, Japan}

\begin{document}\sloppy
\setlength{\abovecaptionskip}{1.0pt}
\setlength{\belowcaptionskip}{0.0pt}
\abovedisplayskip=4pt
\belowdisplayskip=4pt
%\ninept

\maketitle

\begin{abstract}
  In this paper,
  we propose a novel method for separately estimating
  spectral distributions from images captured by a typical RGB camera.
  The proposed method allows us to separately estimate a spectral distribution
  of illumination, reflectance, or camera sensitivity,
  while recent hyperspectral cameras are limited
  to capturing a joint spectral distribution from a scene.
  In addition, the use of Bayesian inference makes it possible
  to take into account prior information of
  both spectral distributions and image noise as probability distributions.
  As a result,
  the proposed method can estimate spectral distributions in a unified way,
  and it can enhance the robustness of the estimation against noise,
  which conventional spectral-distribution estimation methods cannot.
  The use of Bayesian inference also enables us
  to obtain the confidence of estimation results.
  In an experiment,
  the proposed method is shown not only to outperform
  conventional estimation methods in terms of RMSE
  but also to be robust against noise.
\end{abstract}
\renewcommand{\thefootnote}{\fnsymbol{footnote}}
\footnote[0]{This work was supported by JSPS KAKENHI Grant Number JP18J20326.}
\renewcommand{\thefootnote}{\arabic{footnote}}
\section{Introduction}
  Spectral distributions of illumination, the reflectance of objects,
  and camera sensitivity play an important role in color image processing.
  They decide the color of pixel values in images captured by digital cameras.
  Spectral distribution is a function of wavelengths
  and describes what wavelengths of light are included in illumination,
  reflected by objects, or detected by imaging sensors.
  Knowledge on spectral distributions is essential
  for many computer vision tasks that use color information,
  such as hyperspectral imaging~\cite{park2007multispectral,tominaga1999spectral},
  color constancy~\cite{brainard1997bayesian,gijsenij2011computational,cheng2015beyond,afifi2019when,das2018color,afifi2020deep},
  color image enhancement~\cite{ueda2018hue,kinoshita2018automatic_trans,kinoshita2019scene,kinoshita2020hue},
  and intrinsic image decomposition~\cite{cai2017joint,li2018learning,liu2020unsupervised,seo2021deep}.

  Spectral distributions are often measured
  with specialized devices.
  Hyperspectral cameras are available
  for capturing a joint spectral distribution of a scene,
  but the captured distribution is usually not separated into
  illuminance, reflectance, and sensitivity.
  In addition, these cameras require line scans to capture images.
  For separately measuring reflectance or camera sensitivity,
  a monochromator that generates narrow-band light
  and a spectrophotometer that measures the spectral power distribution
  of this light are generally required.
  The requirement of specialized devices,
  including hyperspectral cameras,
  has become a hurdle for estimating spectral distributions
  because these devices are not easy to carry around and are high-cost.

  For this reason, methods of estimating spectral distributions
  with low-cost equipment such as color checkers and commonly used digital cameras
  have been proposed~\cite{han2012camera,jiang2013what,fu2016reflectance}.
  Although Han's and Fu's methods still require a specialized color checker
  or additional light sources,
  Jiang's method~\cite{jiang2013what} can effectively estimate
  camera spectral sensitivity by using only a typical color checker and a camera.
  However, Jiang's method is not available for estimating reflectance
  because it uses an algorithm specifically designed
  for camera sensitivity estimation
  to incorporate prior information of spectral distributions~\cite{judd1964spectral}.
  The simple least-squares method can estimate
  illumination, reflectance, or sensitivity,
  but this method suffers from noise in images.

  Therefore, in this paper,
  we propose a novel method for separately estimating
  a spectral distribution of illumination, reflectance, or camera sensitivity,
  from an image of a color checker captured by a commonly used digital camera.
  The use of Bayesian inference in the method makes it possible to take into account
  prior information of both spectral distributions and image noise
  as probability distributions.
  As a result, the proposed method enables us not only to
  estimate spectral distributions in a unified way
  but also to improve the robustness of the estimation against noise.
  In addition, we can obtain the confidence of the estimation results.

  We performed a numerical simulation to evaluate the performance of the proposed method.
  The method was compared with a simple least-squares-based method
  and Jiang's camera-sensitivity estimation method.
  The experimental results show that the proposed method outperformed
  the two conventional methods
  in terms of the root mean squared error (RMSE)
  between each estimation and the corresponding ground truth.
  The proposed method can also be used to estimate both illumination and sensitivity
  as well as the least-squares method can
  and is demonstrated to be more robust against noise than the least-squares method.
\section{Preliminary}
  In this section, we briefly summarize the imaging pipeline from spectral distributions
  to the pixel values of typical digital cameras.
  After that, we describe problems in estimating spectral distributions
  and clarify our aim.
\subsection{Imaging pipeline}
\begin{figure}[!t]
    \centering
    \includegraphics[width=0.95\columnwidth]{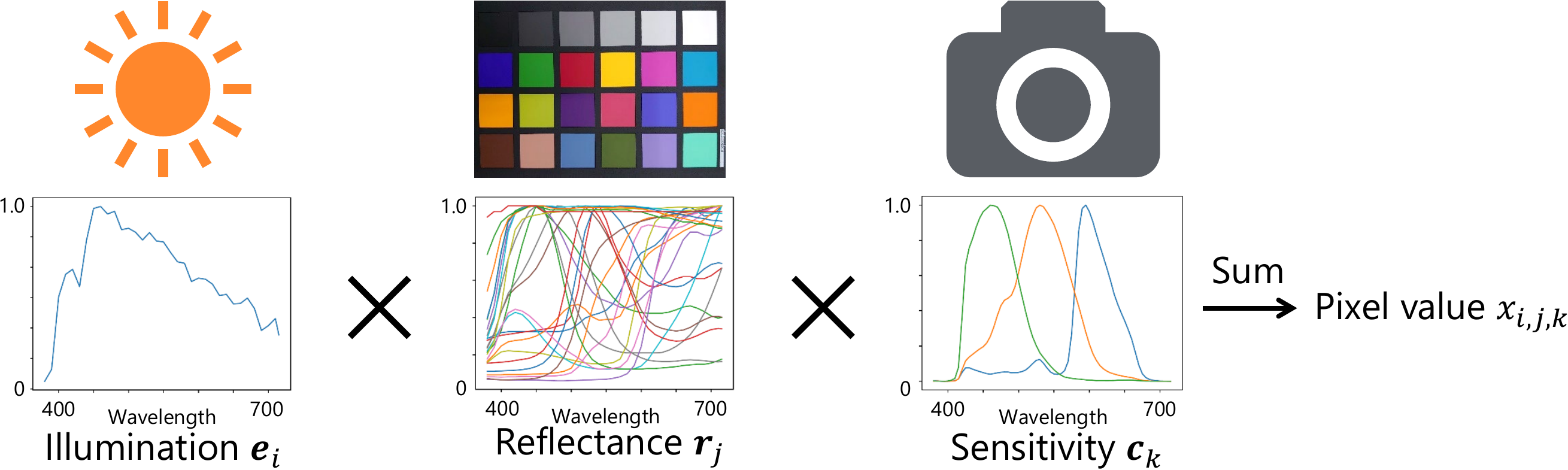}
    \caption{Imaging pipeline of digital camera \label{fig:pipeline}}
  \end{figure}

  Under the assumption that objects have Lambertian reflectance,
  pixel values captured by a digital camera
  can be calculated by using spectral distributions of
  illumination, reflectance, and camera sensitivity
  (see Fig. \ref{fig:pipeline}).
  The mapping from spectral distributions into pixel values $x_{i, j, k}$ is written as:
  \begin{equation}
    x_{i, j, k} = \int_{\mathcal{D}} E_i(\lambda) R_j(\lambda) C_k(\lambda) d\lambda,
    \label{eq:pixel_integral}
  \end{equation}
  where $\lambda$ means a wavelength of light,
  $\mathcal{D}$ is a set of wavelengths of visible light,
  $E_i(\lambda)$ is the $i \in \{1, \cdots, I\}$-th illumination,
  $R_j(\lambda)$ is the $j \in \{1, \cdots, J\}$-th reflectance,
  and $C_k(\lambda)$ is the $k \in \{1, \cdots, K\}$-th sensitivity.
  When we capture an image of a color checker having $4 \times 6$ patches
  by using a single RGB camera under daylight,
  $I, J$, and $K$ are $1, 24$, and $3$, respectively.

  By sampling spectral distributions along the wavelength axis,
  eq. (\ref{eq:pixel_integral}) can be approximated by the following discrete form.
  \begin{equation}
    x_{i, j, k} = \trace{\diag{\myvector{e}_i} \diag{\myvector{r}_j} \diag{\myvector{c}_k}},
    \label{eq:pixel_sum}
  \end{equation}
  where $\myvector{e}_i, \myvector{r}_j$, and $\myvector{c}_k$ are
  $N$-dimensional column vectors
  that denote sampled distributions of $E_i, R_j$, and $C_k$, respectively,
  and $N$ indicates the number of discrete wavelengths.
\subsection{Scenario}
  Our aim in this paper is to separately estimate a spectral distribution of
  illumination, reflectance, or sensitivity
  [i.e., $E_i(\lambda), R_j(\lambda)$, or $C_k(\lambda)$]
  from pixel values $x_{i, j, k}$
  under the condition that the other two spectral distributions are given.
  In contrast, a hyperspectral camera having a large number $K$ of channels
  aims to capture a multiplied spectral distribution $E_i(\lambda)R_j(\lambda)$
  when the camera sensitivity $C_k(\lambda)$ is given.
  
  Let us consider a situation in which we want to estimate illumination $\myvector{e}_i$
  from given reflectance $\myvector{r}_j$, sensitivity $\myvector{c}_k$,
  and the corresponding pixel values $x_{i, j, k}$.
  In this situation, the simplest way is to use least squares as
  \begin{equation}
    \hat{\myvector{e}}_i
    = \argmin_{\myvector{e}}
      \| \myvector{x}_i - \mymatrix{M}_{\myvector{r}, \myvector{{c}}} \myvector{e} \|^2,
    \label{eq:estimation_lsq}
  \end{equation}
  where $\hat{\myvector{e}}_i$ is an estimation of $\myvector{e}_i$.
  Vector $\myvector{x}_i$ is given as
  \begin{equation}
    \myvector{x}_{i} = (x_{i, 1, 1}, x_{i, 1, 2}, \cdots, x_{i, J, K}),
  \end{equation}
  and matrix $\mymatrix{M}_{\myvector{r}, \myvector{c}}$ is calculated
  by using the Hadamard product $\odot$ as
  \begin{equation}
    \mymatrix{M}_{\myvector{r}, \myvector{c}}
      = (\myvector{r}_1 \odot \myvector{c}_1, \myvector{r}_1 \odot \myvector{c}_2,
         \cdots, \myvector{r}_J \odot \myvector{c}_K)^\top.
  \end{equation}
  Similarly to the case of illumination, 
  the least-squares method can also estimate reflectance or sensitivity.
  In practice, however, the direct inversion in eq. (\ref{eq:estimation_lsq})
  cannot reliably recover a spectral distribution
  because it is quite sensitive to noise
  and the spectral reflectance of real-world objects
  has low intrinsic dimensionality, as pointed out in~\cite{jiang2013what}.
  
  To improve the estimation accuracy,
  research has been devoted
  to analyzing the characteristics of spectral distributions
  and to incorporating these characteristics as prior information in the estimation.
  Judd et al. analyzed the spectral distributions of daylight
  and found that they can be written
  as the sum of daylight's mean spectral distribution and
  two other bases~\cite{judd1964spectral}.
  Jiang et al. proposed a camera-spectral-sensitivity estimation method
  using prior information by which camera sensitivity can be approximated
  by using two bases obtained by principal component analysis
  on a set of spectral distributions of
  28 camera sensitivities~\cite{jiang2013what}.
  Jiang's method works better than the least-squares method.
  However, its dedicated algorithm for incorporating
  prior information makes the method specialized
  only for camera sensitivity estimation.
  
  For this reason,
  we propose a novel spectral-distribution estimation method
  based on Bayesian inference.
\section{Proposed method}
  The proposed Bayesian-inference-based spectral-distribution
  estimation method separately estimates the spectral distribution of
  illumination, reflectance, or sensitivity
  from pixel values $x_{i, j, k}$.
  The method enables us to:
  \begin{itemize}[nosep]
    \item Estimate illumination, reflectance, or sensitivity separately in a unified way,
    \item Enhance the robustness of the estimation against noise by 
      incorporating prior information of noise as probability distributions, and
    \item Obtain the confidence of estimation results as a precision matrix.
  \end{itemize}
\subsection{Estimating spectral distribution by Bayesian inference}
  Let us consider a situation in which illumination is estimated
  from given reflectance, sensitivity, and pixel values.
  The goal of the estimation is to calculate the posterior distribution of the
  $\prob{\myvector{e}_i | \myset{X}, \myset{R}, \myset{C}}$ of $\myvector{e}_i$,
  where $\myset{X} = \{x_{i, j, k}\}, \myset{R} = \{\myvector{r}_j\}$,
  and $\myset{C} = \{\myvector{c}_k\}$.
  To estimate $\prob{\myvector{e}_i | \myset{X}, \myset{R}, \myset{C}}$,
  we assume a prior probability distribution $\prob{\myvector{e}_i}$ of $\myvector{e}_i$
  and a probability distribution
  $\prob{x_{i, j, k} | \myvector{e}_i, \myvector{r}_j, \myvector{c}_k}$
  of $x_{i, j, k}$ (i.e., likelihood) as
  \begin{align}
    &\prob{\myvector{e}_i} = \gaussian{\myvector{e}_i}{\myvector{\mu}_{\myvector{e}_i}}{\mymatrix{\Lambda}_{\myvector{e}_i}^{-1}},
      \label{eq:prior_prob}\\
    &\prob{x_{i, j, k} | \myvector{e}_i, \myvector{r}_j, \myvector{c}_k} \nonumber \\
      &= \gaussian{x_{i, j, k}}{\trace{\diag{\myvector{e}_i}\diag{\myvector{r}_j}\diag{\myvector{c}_k}}}{\beta^{-1}},
      \label{eq:likelihood}
  \end{align}
  respectively, where the mean vector $\myvector{\mu}_{\myvector{e}_i}$,
  the precision matrix $\mymatrix{\Lambda}_{\myvector{e}_i}$,
  and the precision $\beta$ are hyperparameters.
  Equation (\ref{eq:likelihood}) corresponds to an image noise model.
  
  On the basis of the Bayesian theorem,
  $\prob{\myvector{e}_i | \myset{X}, \myset{R}, \myset{C}}$ is given by
  \begin{equation}
    \prob{\myvector{e}_i | \myset{X}, \myset{R}, \myset{C}}
      \propto \prob{\myset{X} | \myvector{e}_i, \myset{R}, \myset{C}} \prob{\myvector{e}_i}.
    \label{eq:bayesian_theorem}
  \end{equation}
  Here,
  \begin{equation} 
    \prob{\myset{X} | \myvector{e}_i, \myset{R}, \myset{C}}
      = \prod_{j, k} \prob{x_{i, j, k} | \myvector{e}_i, \myvector{r}_j, \myvector{c}_k}.
  \end{equation}
  Taking the logarithm of $\prob{\myvector{e}_i | \myset{X}, \myset{R}, \myset{C}}$,
  we obtain
  \begin{align}
    &\ln{\prob{\myvector{e}_i | \myset{X}, \myset{R}, \myset{C}}} \nonumber \\
    &= \ln{\prob{\myset{X} | \myvector{e}_i, \myset{R}, \myset{C}}}
      + \ln{\prob{\myvector{e}_i}}
      + \mathrm{const.} \nonumber \\
    &= - \frac{1}{2}
        \myvector{e}_i^\top
        \left(
          \mymatrix{\Lambda}_{\myvector{e}_i}
          + \beta \sum_{j, k} \diag{\myvector{s}_j} \myvector{c}_k \myvector{c}_k^\top \diag{\myvector{s}_j}
        \right)
        \myvector{e}_i \nonumber \\
    &+ \myvector{e}_i^\top
      \left(
        \mymatrix{\Lambda}_{\myvector{e}_i} \myvector{\mu}_{\myvector{e}_i} + \beta \sum_{j, k} x_{i, j, k} \diag{\myvector{s}_j} \myvector{c}_k
      \right) \nonumber \\
    &+ \mathrm{const.}
    \label{eq:log_prob}
  \end{align}
  From eq. (\ref{eq:log_prob}),
  it is confirmed that the posterior distribution $\prob{\myvector{e}_i | \myset{X}, \myset{R}, \myset{C}}$
  follows a Gaussian distribution
  $\gaussian{\myvector{e}_i}{\hat{\myvector{\mu}}_{\myvector{e}_i}}{\hat{\mymatrix{\Lambda}}_{\myvector{e}_i}}$.
  The mean $\hat{\myvector{\mu}}_{\myvector{e}_i}$ and precision $\hat{\mymatrix{\Lambda}}_{\myvector{e}_i}$
  of the posterior distribution are given as
  \begin{align}
    \hat{\mymatrix{\Lambda}}_{\myvector{e}_i}
      &= \mymatrix{\Lambda}_{\myvector{e}_i}
      - \beta \sum_{j, k} \diag{\myvector{r}_j} \myvector{c}_k \myvector{c}_k^\top \diag{\myvector{r}_j},
      \label{eq:posterior_precision_illumination}\\
    \hat{\myvector{\mu}}_{\myvector{e}_i}
      &= \hat{\mymatrix{\Lambda}}_{\myvector{e}_i}
        \left(
          \Lambda_{\myvector{e}_i} \myvector{\mu}_{\myvector{e}_i}
          + \beta \sum_{j, k} x_{i, j, k} \diag{\myvector{r}_j} \myvector{c}_k
        \right).
    \label{eq:posterior_mean_illumination}
  \end{align}
  
  Similarly to the case of illumination,
  we can estimate
  the posterior probability distribution $\prob{\myvector{r}_j | \myset{X}, \myset{E}, \myset{C}}$
  of reflectance from illumination, sensitivity, and pixel values as
  \begin{align}
    \hat{\mymatrix{\Lambda}}_{\myvector{r}_j}
      &= \mymatrix{\Lambda}_{\myvector{r}_j}
      - \beta \sum_{k, i} \diag{\myvector{c}_k} \myvector{e}_i \myvector{e}_i^\top \diag{\myvector{c}_k},
      \label{eq:posterior_precision_reflectance}\\
    \hat{\myvector{\mu}}_{\myvector{r}_j}
      &= \hat{\mymatrix{\Lambda}}_{\myvector{r}_j}
        \left(
          \mymatrix{\Lambda}_{\myvector{r}_j} \myvector{\mu}_{\myvector{r}_j}
          + \beta \sum_{k, i} x_{i, j, k} \diag{\myvector{c}_k} \myvector{e}_i
        \right),
    \label{eq:posterior_mean_reflectance}
  \end{align}
  or we can estimate
  the posterior probability distribution $\prob{\myvector{c}_k | \myset{X}, \myset{E}, \myset{R}}$
  of sensitivity from illumination, reflectance, and pixel values as
  \begin{align}
    \hat{\mymatrix{\Lambda}}_{\myvector{c}_k}
      &= \mymatrix{\Lambda}_{\myvector{c}_k}
      - \beta \sum_{i, j} \diag{\myvector{e}_i} \myvector{r}_j \myvector{r}_j^\top \diag{\myvector{e}_i},
      \label{eq:posterior_precision_sensitivity}\\
    \hat{\myvector{\mu}}_{\myvector{c}_k}
      &= \hat{\mymatrix{\Lambda}}_{\myvector{c}_k}
        \left(
          \mymatrix{\Lambda}_{\myvector{c}_k} \myvector{\mu}_{\myvector{c}_k}
          + \beta \sum_{i, j} x_{i, j, k} \diag{\myvector{e}_i} \myvector{r}_j
        \right).
    \label{eq:posterior_mean_sensitivity}
  \end{align}

\subsection{Incorporating prior information into estimation}
  Incorporating prior information of spectral distributions
  into the proposed estimation method is simply done
  by modifying the mean and the precision matrix for the prior distributions
  $\prob{\myvector{e}_i}, \prob{\myvector{r}_j}$, and $\prob{\myvector{c}_k}$.
  For example,
  we can use the mean spectral distribution $\overline{\myvector{e}}$
  of daylight in~\cite{judd1964spectral} as 
  \begin{equation}
    \myvector{\mu}_{\myvector{e}_i} = \overline{\myvector{e}}.
    \label{eq:daylight_prior}
  \end{equation}
  In addition,
  the mean spectral distribution $\overline{\myvector{c}}_k$
  of the $28$ RGB camera sensitivities in~\cite{judd1964spectral}
  can be utilized as 
  \begin{equation}
    \myvector{\mu}_{\myvector{c}_k} = \overline{\myvector{c}}_k.
    \label{eq:sensitivity_prior}
  \end{equation}

  Incorporating prior information of image noise
  into the proposed estimation method is similarly done
  by modifying the mean and the precision for
  $\prob{x_{i, j, k} | \myvector{e}_i, \myvector{r}_j, \myvector{c}_k}$.
\subsection{Proposed procedure}
  The procedure for estimating illumination $\myvector{e}_i$
  by using the proposed method is as follows
  \begin{enumerate}[nosep]
    \item Set hyperparameters for prior distributions
      in eqs. (\ref{eq:prior_prob}) and (\ref{eq:likelihood}),
      where prior information of illumination can be incorporated
      such as in eq. (\ref{eq:daylight_prior}).
    \item Calculate the mean $\hat{\myvector{\mu}}_{\myvector{e}_i}$
      and the precision matrix $\hat{\mymatrix{\Lambda}}_{\myvector{e}_i}$
      of the posterior distribution
      $\prob{\myvector{e}_i | \myset{X}, \myset{R}, \myset{C}}$
      in accordance with eqs. (\ref{eq:posterior_precision_illumination})
      and (\ref{eq:posterior_mean_illumination})
      by using $\myset{X}, \myset{R}$, and $\myset{C}$.
    \item Obtain an estimation
      by normalizing the mean $\hat{\myvector{\mu}}_{\myvector{e}_i}$ as
      \begin{equation}
        \hat{\myvector{e}}_i
          = \frac{\hat{\myvector{\mu}}_{\myvector{e}_i}}{\underset{n}{\max}~e_{i, n}},
        \label{eq:normalize}
      \end{equation}
      where $e_{i, n}$ indicates the $n$-th element of $\myvector{e}_i$.
  \end{enumerate}
  Note that the calculated posterior distribution
  $\prob{\myvector{e}_i | \myset{X}, \myset{R}, \myset{C}}$
  can be used as the prior distribution $\prob{\myvector{e}_i}$ for the next estimation.
  This enables the estimation to be updated each time a new observation (i.e., captured image)
  is obtained.
  In addition, $\hat{\mymatrix{\Lambda}}_{\myvector{e}_i}$ can be considered as
  the confidence of the estimation.
  Reflectance $\myvector{r}_j$ and sensitivity $\myvector{c}_k$
  can be estimated in the same way as for illumination
  [see eqs. (\ref{eq:posterior_precision_illumination}) to (\ref{eq:posterior_mean_sensitivity})].
\section{Simulation}
  To confirm the effectiveness of the proposed method,
  we conducted a simulation.
\subsection{Simulation conditions}
\begin{figure}[!t]
    \centering
    \begin{subfigure}[!t]{0.31\columnwidth}
      \centering
      \includegraphics[width=\columnwidth]{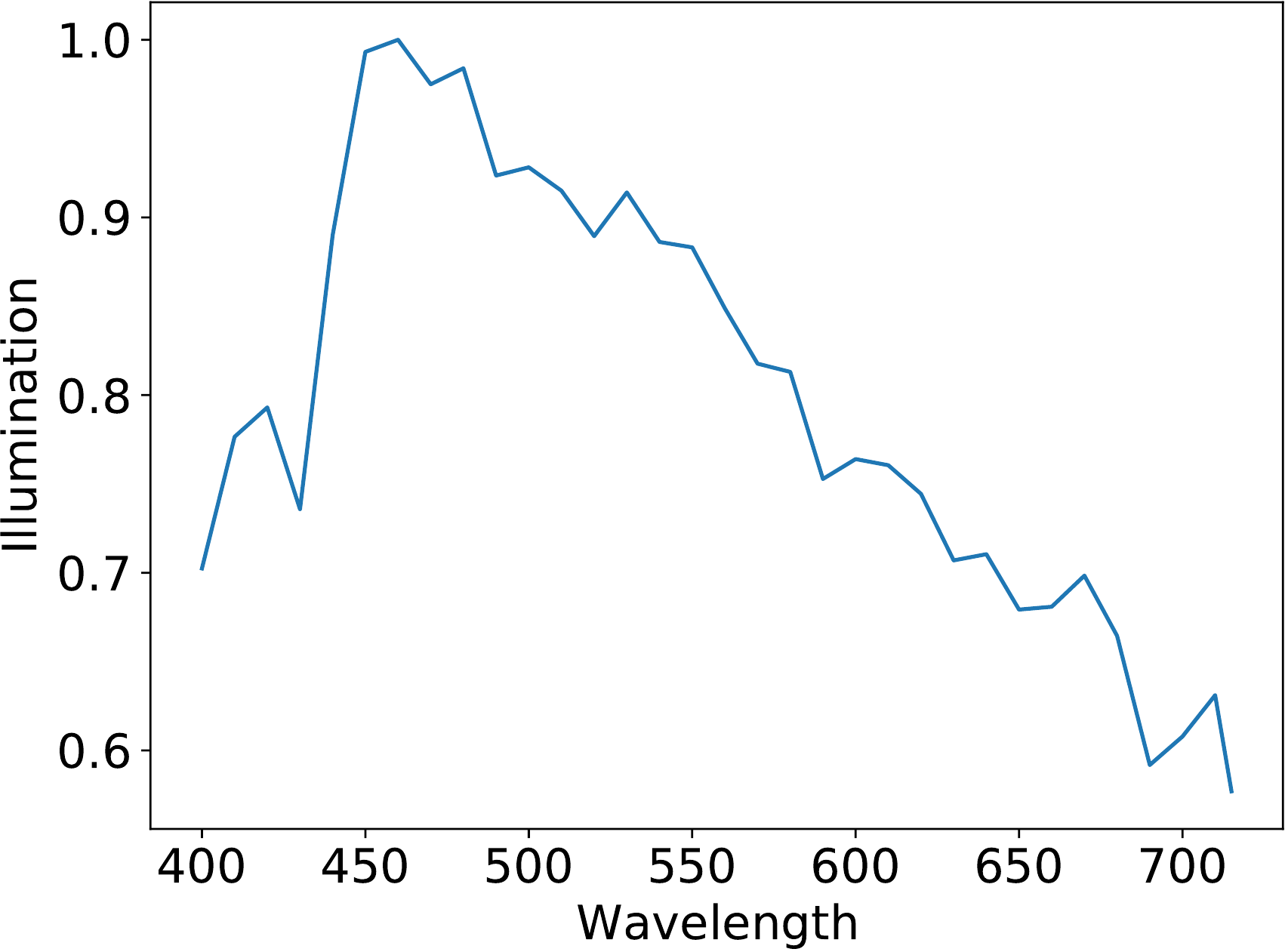}
      \caption{Illumination \label{fig:illumination}}
    \end{subfigure}
    \begin{subfigure}[!t]{0.31\columnwidth}
      \centering
      \includegraphics[width=\columnwidth]{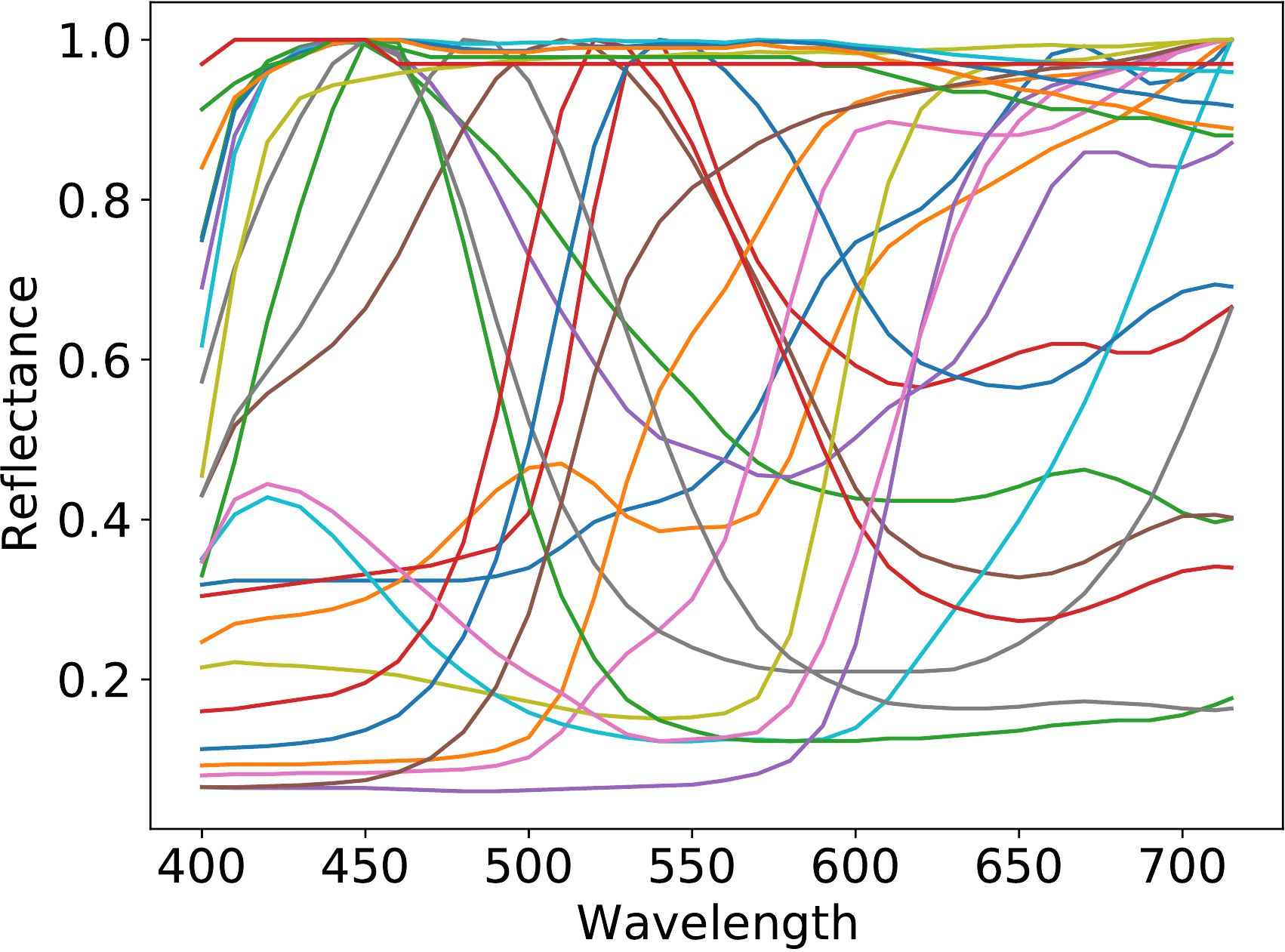}
      \caption{Reflectance \label{fig:reflectance}}
    \end{subfigure}
    \begin{subfigure}[!t]{0.31\columnwidth}
      \centering
      \includegraphics[width=\columnwidth]{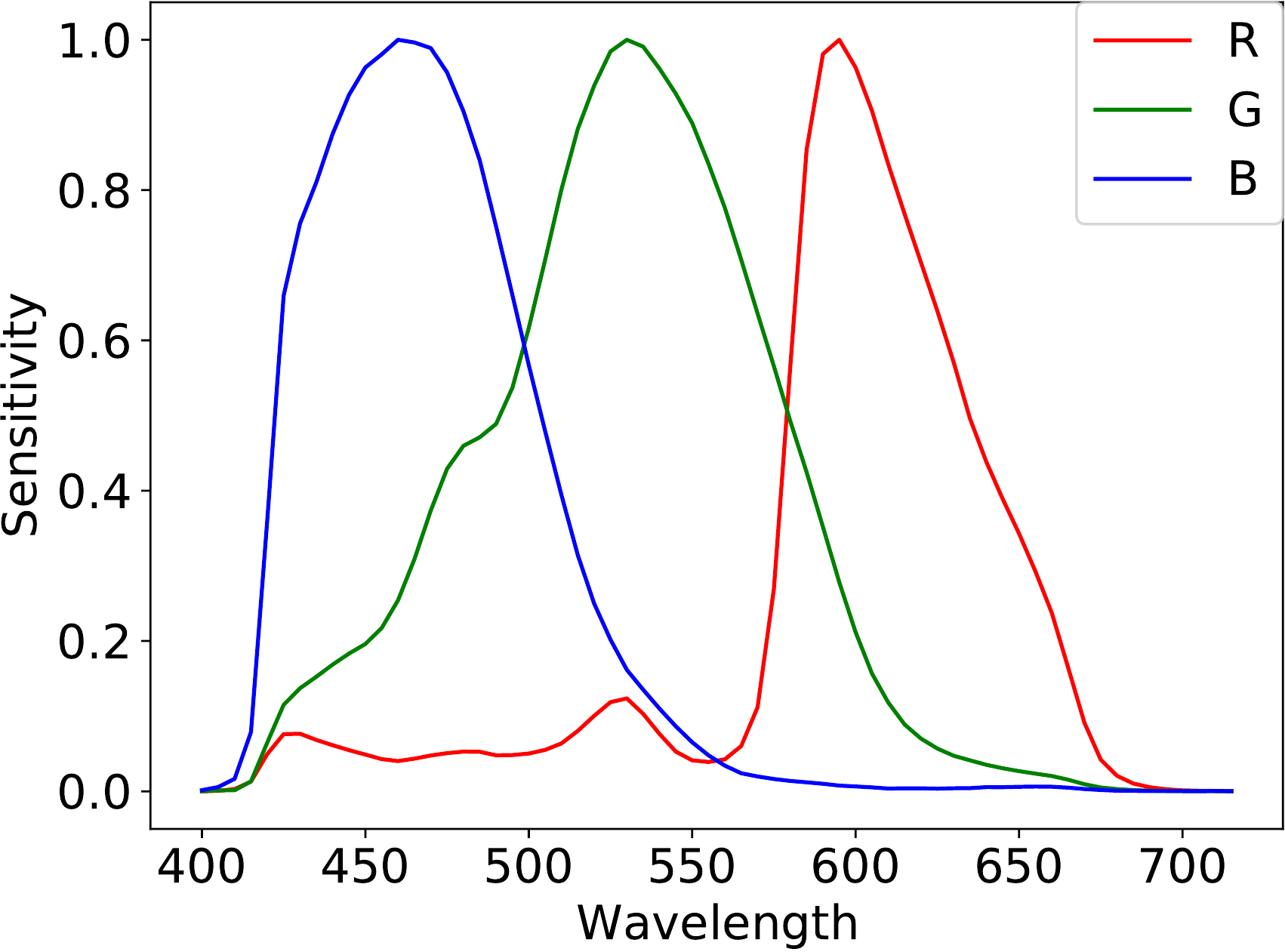}
      \caption{Sensitivity \label{fig:sensitivity}}
    \end{subfigure}
    \caption{Ground truth for spectral distribution \label{fig:groundtruth}}
  \end{figure}

  In this simulation,
  we calculated pixel values from the spectral distributions shown in Fig. \ref{fig:groundtruth}
  in accordance with eq. (\ref{eq:pixel_sum}),
  where the illumination was from the spectral distribution of the CIE Standard Illuminant D65,
  the reflectance was from a color checker manufactured by BabelColor~\cite{babelcolor},
  and the camera sensitivity was that of the Nikon 5100 in~\cite{colour}.
  In addition,
  we added Gaussian noise with a mean of 0 and a standard deviation of $0.01$
  into the calculated pixel values $x_{i, j, k}$.
  Assuming a situation in which the illumination is unknown,
  we estimated the illumination by using the reflectance, sensitivity, and pixel values.
  Similarly to the illumination,
  we also estimated the sensitivity from the illumination, reflectance, and pixel values.
  The estimation accuracy of the proposed method was evaluated by using the well-known
  root mean squared error (RMSE) between the estimation and the corresponding ground truth.

  Although there are advanced spectral-distribution estimation methods
  such as~\cite{han2012camera,fu2016reflectance},
  they require a specialized color checker or light source.
  For this reason, 
  we compared the proposed method with
  the least-squares method shown in eq. (\ref{eq:estimation_lsq})
  and with Jiang's high-performance method that requires no special equipment
  ~\cite{jiang2013what}.
  Here, we used Jiang's method only for estimating sensitivity
  since Jiang's method is not designed for estimating illumination
  when sensitivity is given.
\subsection{Results}
\begin{figure}[!t]
    \centering
    \begin{subfigure}[!t]{0.47\columnwidth}
      \centering
      \includegraphics[width=\columnwidth]{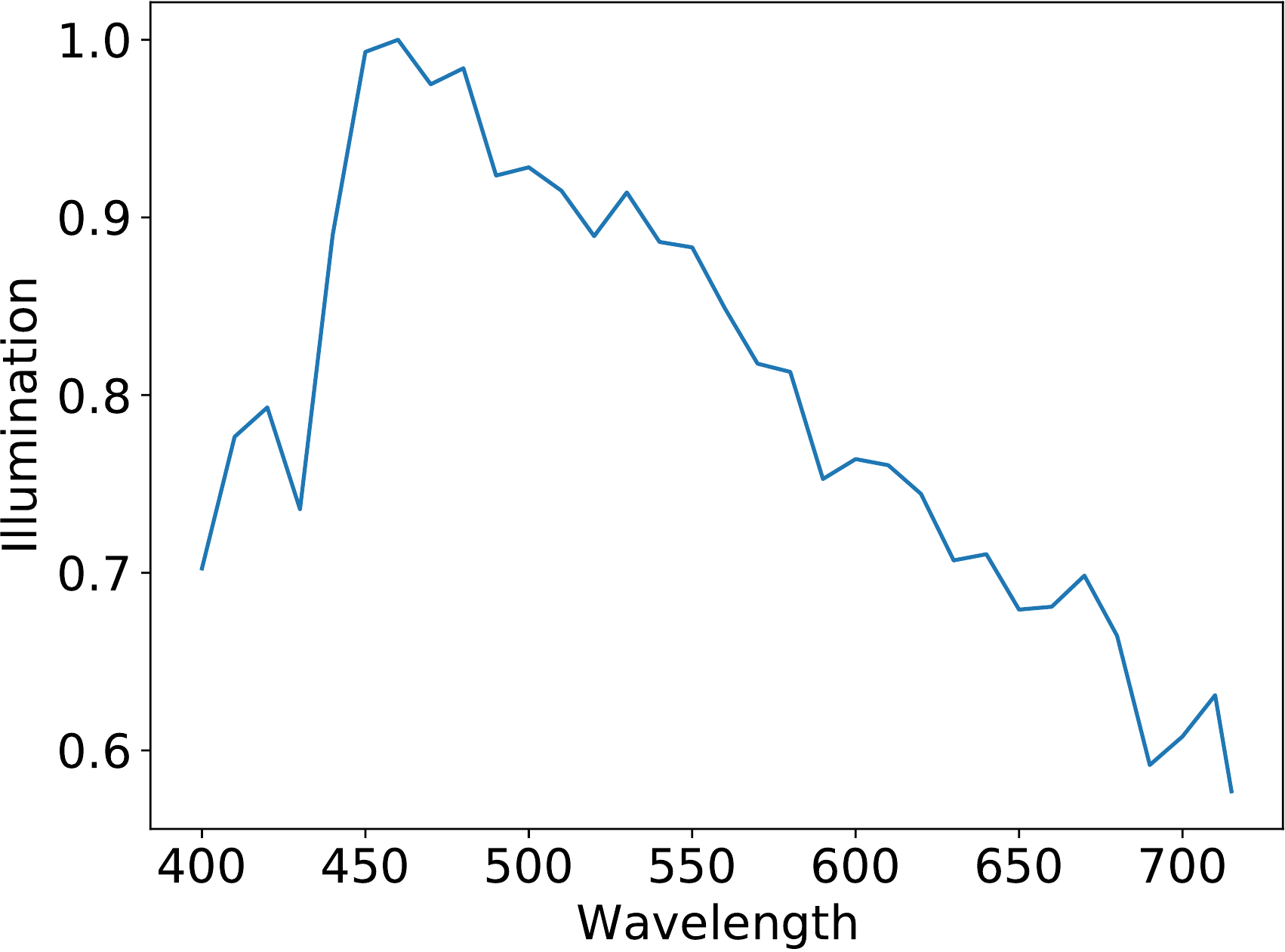}
      \caption{Least squares \label{fig:lsq_illumination}}
    \end{subfigure}
    \begin{subfigure}[!t]{0.47\columnwidth}
      \centering
      \includegraphics[width=\columnwidth]{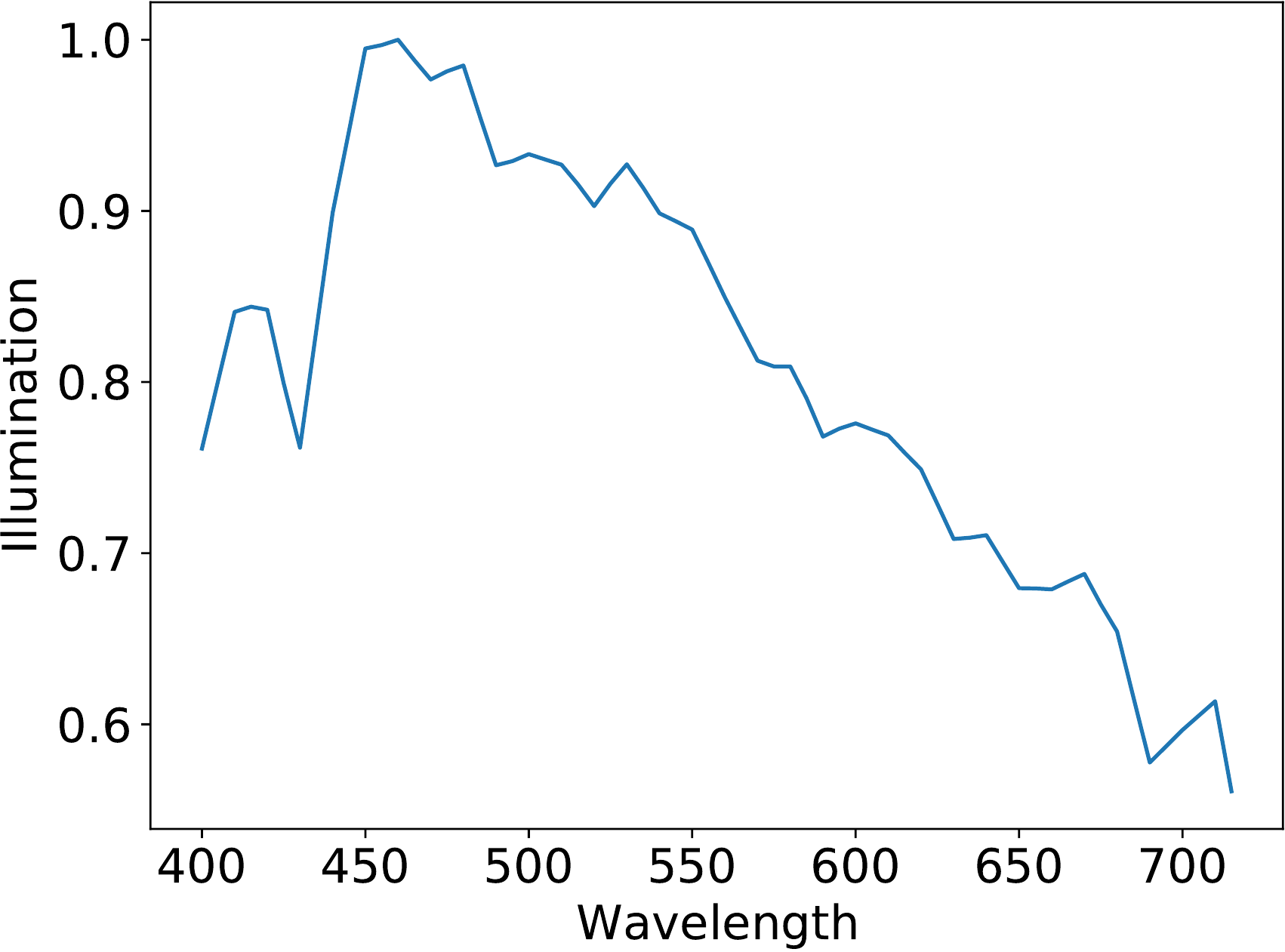}
      \caption{Proposed \label{fig:prop_illumination}}
    \end{subfigure}
    \caption{Estimated illumination \label{fig:estimated_illumination}}
  \end{figure}

  Figure \ref{fig:estimated_illumination} shows
  spectral distributions of the illumination estimated by using the least-squares
  and proposed methods in the case where pixel values had no noise.
  Both methods estimated the illumination well.

\begin{figure}[!t]
    \centering
    \begin{subfigure}[!t]{0.31\columnwidth}
      \centering
      \includegraphics[width=\columnwidth]{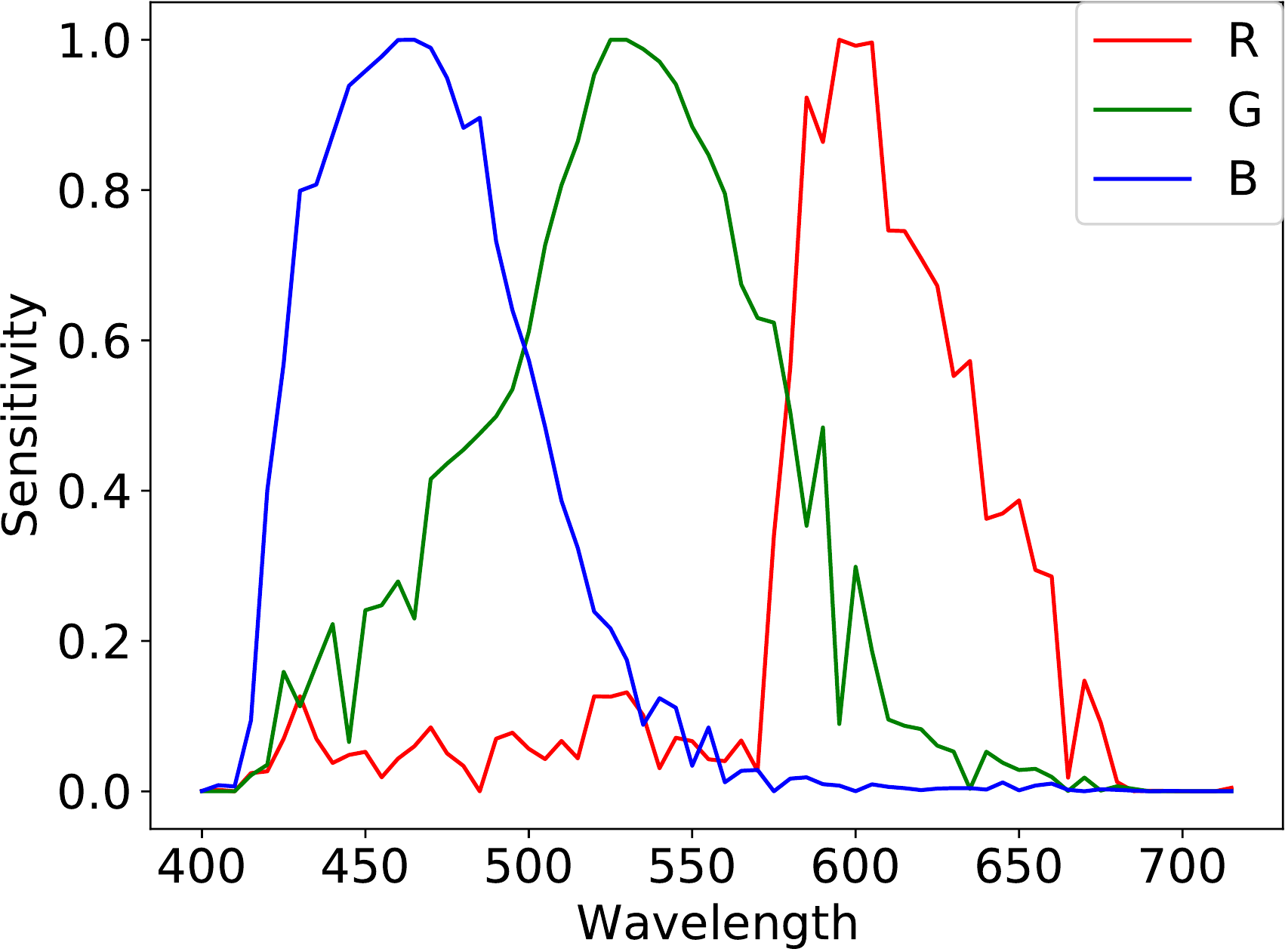}
      \caption{Least squares \label{fig:lsq_sensitivity}}
    \end{subfigure}
    \begin{subfigure}[!t]{0.31\columnwidth}
      \centering
      \includegraphics[width=\columnwidth]{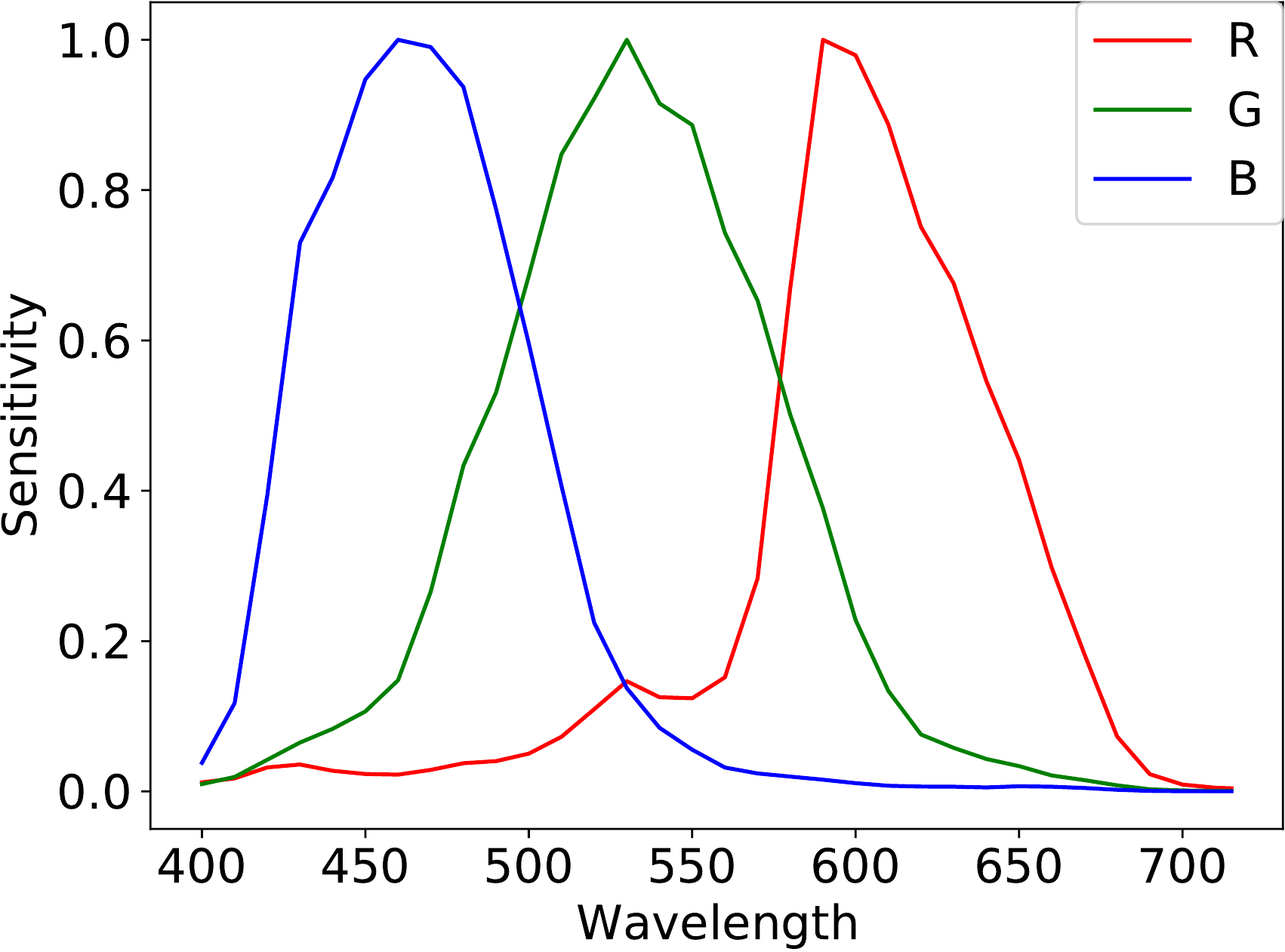}
      \caption{Jiang~\cite{jiang2013what} \label{fig:jiang_sensitivity}}
    \end{subfigure}
    \begin{subfigure}[!t]{0.31\columnwidth}
      \centering
      \includegraphics[width=\columnwidth]{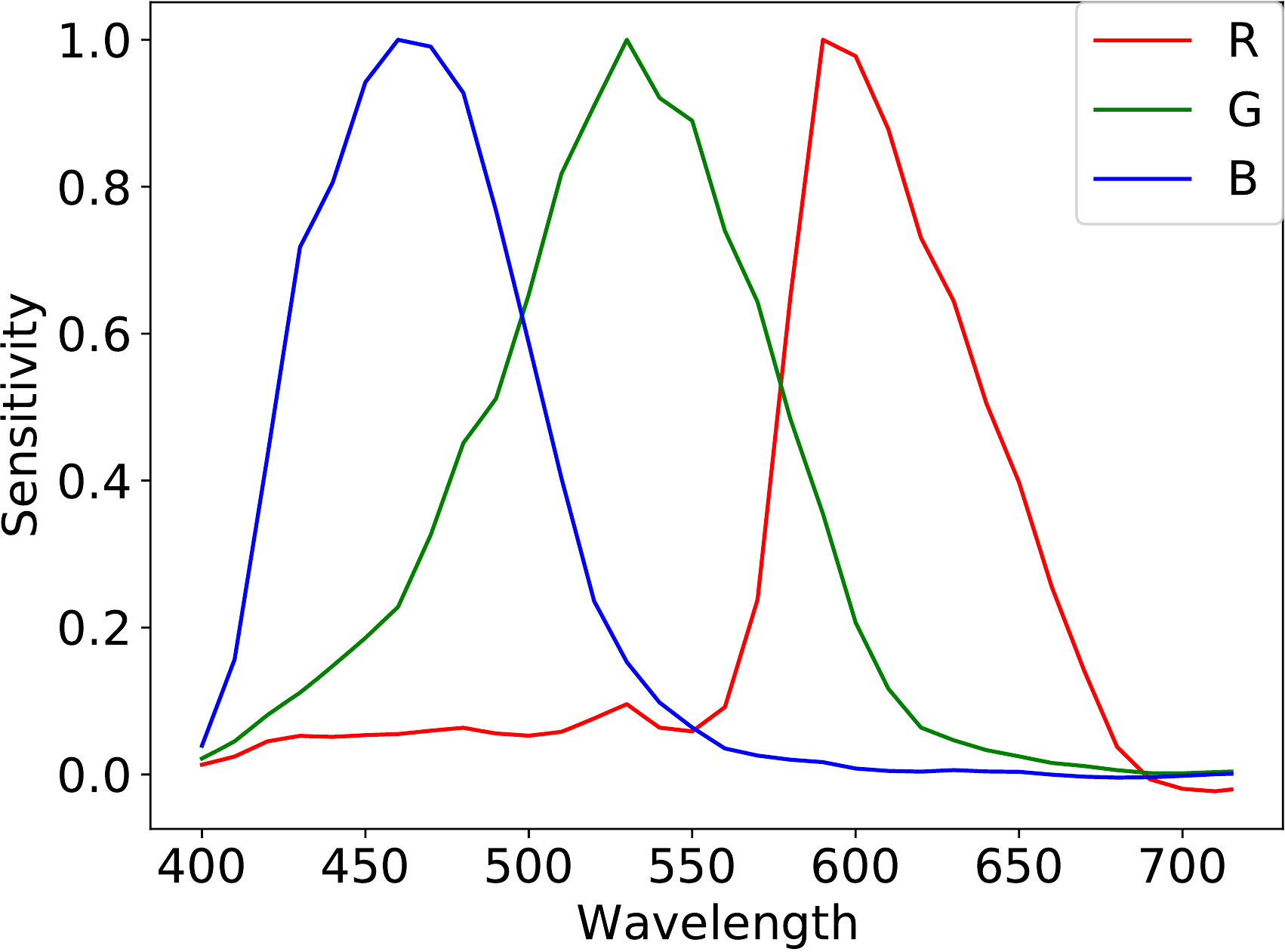}
      \caption{Proposed \label{fig:prop_sensitivity}}
    \end{subfigure}
    \caption{Estimated sensitivity \label{fig:estimated_sensitivity}}
  \end{figure}

  Figure \ref{fig:estimated_sensitivity} shows
  spectral distributions of the sensitivity estimated by using the three methods
  for the case that pixel values had no noise.
  As shown in Fig. \ref{fig:estimated_sensitivity}(\subref{fig:lsq_sensitivity}),
  least squares provided an estimation that was not smooth even under the noiseless case,
  but Jiang's method and the proposed method provided smooth distributions.
  This is because the number of available pixel values for estimating sensitivity
  was less than that for estimating illumination,
  i.e., $I \times J < J \times K$ in this case.
  
  RMSE scores for the three methods in the case of no noise and noise
  with the standard deviation of $0.01$
  are shown in Tables \ref{tab:result_nonoise} and \ref{tab:result_with_noise},
  respectively.
  From the tables,
  it is confirmed that the proposed method and Jiang's method were robust against noise,
  but the least-squares method was not.
  \begin{table}[!t]
    \caption{RMSE scores (w/o noise) \label{tab:result_nonoise}}
    \begin{tabular}{l|ccc} \hline \hline
                    & Least squares  & Jiang~\cite{jiang2013what} & Proposed       \\ \hline
      Illumination  & \textbf{0.000} & ---   & 0.019          \\
      Sensitivity R & \textbf{0.041} & 0.064 & 0.043          \\
      Sensitivity G & 0.041          & 0.043 & \textbf{0.020} \\
      Sensitivity B & \textbf{0.019} & 0.034 & 0.040          \\ \hline
    \end{tabular}
  \end{table}
  \begin{table}[!t]
    \caption{RMSE scores (w/ additive Gaussian noise having std. of 0.01) \label{tab:result_with_noise}}
    \begin{tabular}{l|ccc} \hline \hline
                    & Least squares  & Jiang~\cite{jiang2013what} & Proposed       \\ \hline
      Illumination  & 0.580         & ---            & \textbf{0.018} \\
      Sensitivity R & 0.149         & 0.064          & \textbf{0.043} \\
      Sensitivity G & 0.256         & 0.042          & \textbf{0.020} \\
      Sensitivity B & 0.168         & \textbf{0.034} & 0.040  \\ \hline
    \end{tabular}
  \end{table}

  For these reasons,
  the effectiveness of the proposed method was confirmed in terms of RMSE.
  The proposed method was also demonstrated to be robust against noise
  and to achieve accuracy comparable to Jiang's recent effective method.
  Furthermore,
  the proposed method can estimate illumination, reflectance,
  or sensitivity in a unified way,
  while Jiang's method is specialized only for estimating sensitivity.
\section{Conclusion}
  In this paper,
  we proposed a novel method for separately estimating a spectral distribution
  of illumination, reflectance, or sensitivity
  from images of a color checker captured by a single RGB camera.
  By using Bayesian inference in the method,
  prior information of both spectral distributions and image noise
  can be incorporated into the estimation.
  As a result, the method enables us not only to
  estimate spectral distributions in a unified way
  but also to improve the robustness of the estimation against noise.
  The method also has other advantages of Bayesian inference,
  i.e., obtaining the confidence of estimation.
  Experimental results showed that
  the proposed method can correctly estimate spectral distributions
  even under a noisy condition.

  In future work,
  we will extend the proposed method to estimate both illumination and sensitivity
  at the same time from pixel values and reflectance.
  Moreover, we will apply a more realistic noise model to the proposed method.
%
%\bibliographystyle{IEEEtran}
%\bibliography{../../../../../bibliography/english_papers,../../../../../bibliography/standards,../../../../../bibliography/web_pages}

\begin{thebibliography}{10}
\providecommand{\url}[1]{#1}
\csname url@samestyle\endcsname
\providecommand{\newblock}{\relax}
\providecommand{\bibinfo}[2]{#2}
\providecommand{\BIBentrySTDinterwordspacing}{\spaceskip=0pt\relax}
\providecommand{\BIBentryALTinterwordstretchfactor}{4}
\providecommand{\BIBentryALTinterwordspacing}{\spaceskip=\fontdimen2\font plus
\BIBentryALTinterwordstretchfactor\fontdimen3\font minus
  \fontdimen4\font\relax}
\providecommand{\BIBforeignlanguage}[2]{{%
\expandafter\ifx\csname l@#1\endcsname\relax
\typeout{** WARNING: IEEEtran.bst: No hyphenation pattern has been}%
\typeout{** loaded for the language `#1'. Using the pattern for}%
\typeout{** the default language instead.}%
\else
\language=\csname l@#1\endcsname
\fi
#2}}
\providecommand{\BIBdecl}{\relax}
\BIBdecl

\bibitem{park2007multispectral}
\BIBentryALTinterwordspacing
J.-I. Park, M.-H. Lee, M.~D. Grossberg, and S.~K. Nayar, ``{Multispectral
  Imaging Using Multiplexed Illumination},'' in \emph{Proceedings of IEEE
  International Conference on Computer Vision},
  Oct. 2007, pp. 1--8.
\BIBentrySTDinterwordspacing

\bibitem{tominaga1999spectral}
\BIBentryALTinterwordspacing
S.~Tominaga, ``{Spectral imaging by a multichannel camera},'' \emph{Journal of
  Electronic Imaging}, vol.~8, no.~4, p. 332, Oct. 1999.
\BIBentrySTDinterwordspacing

\bibitem{brainard1997bayesian}
D.~H. Brainard and W.~T. Freeman, ``{Bayesian color constancy},'' \emph{Journal
  of the Optical Society of America A}, vol.~14, no.~7, pp. 1393--1411, 1997.

\bibitem{gijsenij2011computational}
\BIBentryALTinterwordspacing
A.~Gijsenij, T.~Gevers, and J.~van~de Weijer, ``{Computational Color Constancy:
  Survey and Experiments},'' \emph{IEEE Transactions on Image Processing},
  vol.~20, no.~9, pp. 2475--2489, Sep. 2011.
\BIBentrySTDinterwordspacing

\bibitem{cheng2015beyond}
\BIBentryALTinterwordspacing
D.~Cheng, B.~Price, S.~Cohen, and M.~S. Brown, ``{Beyond White: Ground Truth
  Colors for Color Constancy Correction},'' in \emph{Proceedings of IEEE
  International Conference on Computer Vision},
  Dec. 2015, pp. 298--306.
\BIBentrySTDinterwordspacing

\bibitem{afifi2019when}
\BIBentryALTinterwordspacing
M.~Afifi, B.~Price, S.~Cohen, and M.~S. Brown, ``{When Color Constancy Goes
  Wrong: Correcting Improperly White-Balanced Images},'' in \emph{Proceedings
  of IEEE Conference on Computer Vision and Pattern Recognition},
  Jun. 2019, pp. 1535--1544.
\BIBentrySTDinterwordspacing

\bibitem{das2018color}
\BIBentryALTinterwordspacing
P.~Das, A.~S. Baslamisli, Y.~Liu, S.~Karaoglu, and T.~Gevers, ``{Color
  Constancy by GANs: An Experimental Survey},'' Dec. 2018. [Online]. Available:
  \url{http://arxiv.org/abs/1812.03085}
\BIBentrySTDinterwordspacing

\bibitem{afifi2020deep}
\BIBentryALTinterwordspacing
M.~Afifi and M.~S. Brown, ``{Deep White-Balance Editing},'' in
  \emph{Proceedings of IEEE Conference on Computer Vision and Pattern
  Recognition}, Jun. 2020, pp. 1394--1403.
\BIBentrySTDinterwordspacing

\bibitem{ueda2018hue}
\BIBentryALTinterwordspacing
Y.~Ueda, H.~Misawa, T.~Koga, N.~Suetake, and E.~Uchino, ``{Hue-Preserving Color
  Contrast Enhancement Method Without Gamut Problem by Using Histogram
  Specification},'' in \emph{Proceedings of IEEE International Conference on
  Image Processing}, Oct.
  2018, pp. 1123--1127.
\BIBentrySTDinterwordspacing

\bibitem{kinoshita2018automatic_trans}
\BIBentryALTinterwordspacing
Y.~Kinoshita and H.~Kiya, ``{Automatic exposure compensation using an image
  segmentation method for single-image-based multi-exposure fusion},''
  \emph{APSIPA Transactions on Signal and Information Processing}, vol.~7, no.
  e22, Dec. 2018.
\BIBentrySTDinterwordspacing

\bibitem{kinoshita2019scene}
\BIBentryALTinterwordspacing
Y.~Kinoshita and H.~Kiya, ``{Scene Segmentation-Based Luminance Adjustment for Multi-Exposure
  Image Fusion},'' \emph{IEEE Transactions on Image Processing}, vol.~28,
  no.~8, pp. 4101--4116, Aug. 2019.
\BIBentrySTDinterwordspacing

\bibitem{kinoshita2020hue}
\BIBentryALTinterwordspacing
Y.~Kinoshita and H.~Kiya, ``{Hue-Correction Scheme Based on Constant-Hue Plane
  for Deep-Learning-Based Color-Image Enhancement},'' \emph{IEEE Access},
  vol.~8, pp. 9540--9550, Jan. 2020.
\BIBentrySTDinterwordspacing

\bibitem{cai2017joint}
\BIBentryALTinterwordspacing
B.~Cai, X.~Xu, K.~Guo, K.~Jia, B.~Hu, and D.~Tao, ``{A Joint
  Intrinsic-Extrinsic Prior Model for Retinex},'' in \emph{Proceedings of IEEE
  International Conference on Computer Vision},
  Oct. 2017, pp. 4020--4029.
\BIBentrySTDinterwordspacing

\bibitem{li2018learning}
Z.~Li and N.~Snavely, ``{Learning Intrinsic Image Decomposition from Watching
  the World},'' in \emph{Proceedings of IEEE Conference on Computer Vision and
  Pattern Recognition},
  Jun. 2018, pp. 9039--9048.

\bibitem{liu2020unsupervised}
\BIBentryALTinterwordspacing
Y.~Liu, Y.~Li, S.~You, and F.~Lu, ``{Unsupervised Learning for Intrinsic Image
  Decomposition from a Single Image},'' in \emph{Proceedings of IEEE Conference
  on Computer Vision and Pattern Recognition}, Jun. 2020, pp.
  3245--3254.
\BIBentrySTDinterwordspacing

\bibitem{seo2021deep}
\BIBentryALTinterwordspacing
K.~Seo, Y.~Kinoshita, and H.~Kiya, ``{Deep Retinex Network for Estimating
  Illumination Colors with Self-Supervised Learning},'' in \emph{Proceedings of
  IEEE Global Conference on Life Sciences and Technologies}, Mar. 2021, pp.
  1--5.
\BIBentrySTDinterwordspacing

\bibitem{han2012camera}
\BIBentryALTinterwordspacing
{Shuai Han}, Y.~Matsushita, I.~Sato, T.~Okabe, and Y.~Sato, ``{Camera spectral
  sensitivity estimation from a single image under unknown illumination by
  using fluorescence},'' in \emph{Proceedings of IEEE Conference on Computer
  Vision and Pattern Recognition},
  Jun. 2012, pp. 805--812.
\BIBentrySTDinterwordspacing

\bibitem{fu2016reflectance}
\BIBentryALTinterwordspacing
Y.~Fu, A.~Lam, I.~Sato, T.~Okabe, and Y.~Sato, ``{Reflectance and Fluorescence
  Spectral Recovery via Actively Lit RGB Images},'' \emph{IEEE Transactions on
  Pattern Analysis and Machine Intelligence}, vol.~38, no.~7, pp. 1313--1326,
  Jul. 2016.
\BIBentrySTDinterwordspacing

\bibitem{jiang2013what}
\BIBentryALTinterwordspacing
J.~Jiang, D.~Liu, J.~Gu, and S.~Susstrunk, ``{What is the space of spectral
  sensitivity functions for digital color cameras?}'' in \emph{Proceedings of
  IEEE Workshop on Applications of Computer Vision},
  Jan. 2013, pp. 168--179.
\BIBentrySTDinterwordspacing

\bibitem{judd1964spectral}
\BIBentryALTinterwordspacing
D.~B. Judd, D.~L. MacAdam, G.~Wyszecki, H.~W. Budde, H.~R. Condit, S.~T.
  Henderson, and J.~L. Simonds, ``{Spectral Distribution of Typical Daylight as
  a Function of Correlated Color Temperature},'' \emph{Journal of the Optical
  Society of America}, vol.~54, no.~8, p. 1031, Aug. 1964.
\BIBentrySTDinterwordspacing

\bibitem{babelcolor}
\BIBentryALTinterwordspacing
``{BabelColor}.'' [Online]. Available:
  \url{https://www.babelcolor.com/colorchecker-2.htm}
\BIBentrySTDinterwordspacing

\bibitem{colour}
\BIBentryALTinterwordspacing
T.~Mansencal, M.~Mauderer, M.~Parsons, N.~Shaw, K.~Wheatley, S.~Cooper, J.~D.
  Vandenberg, L.~Canavan, K.~Crowson, O.~Lev, K.~Leinweber, S.~Sharma, T.~J.
  Sobotka, D.~Moritz, M.~Pppp, C.~Rane, P.~Eswaramoorthy, J.~Mertic,
  B.~Pearlstine, M.~Leonhardt, O.~Niemitalo, M.~Szymanski, M.~Schambach,
  S.~Huang, M.~Wei, N.~Joywardhan, O.~Wagih, P.~Redman, J.~Goldstone, and
  S.~Hill, ``{Colour 0.3.16},''
  [Online]. Available:
  \url{https://doi.org/10.5281/zenodo.3757045#.X_zJFXCo-EY.mendeley}
\BIBentrySTDinterwordspacing

\end{thebibliography}

\end{document}